\documentclass[prl,superscriptaddress, letterpaper, fleqn, floatfix, showpacs, twocolumn]{revtex4}
\usepackage{euscript, units, amsfonts,amsmath, graphics, graphicx, dcolumn, fancyhdr}
\usepackage{times}
\usepackage{sidecap}

\usepackage[usenames]{color}

\begin{document}
\title{Hou \emph{et al.} Reply to the comment on ``Self-Diffusion in 2D Dusty-Plasma Liquids: Numerical-Simulation Results''}\author{Lu-Jing Hou}
\affiliation{Institut f\"{u}r Experimentelle und Angewandte Physik, Christian-Albrechts Universit\"{a}t zu Kiel, 
D-24098 Kiel, Germany}
\author{Alexander Piel}
\affiliation{Institut f\"{u}r Experimentelle und Angewandte Physik, Christian-Albrechts Universit\"{a}t zu Kiel, 
D-24098 Kiel, Germany}
\author{P. K. Shukla}
\affiliation{Institut f\"{u}r Theoretische Physik IV, Ruhr-Universit\"{a}t Bochum, D-44780, Germany}
\begin{abstract}
Reply to the comment on ``Self-Diffusion in 2D Dusty-Plasma Liquids: Numerical-Simulation Results''\end{abstract}

 \maketitle

\noindent 
The preceding Comment \cite{Comment} makes a few claims on the inserted figure of Fig. 5 in our Letter \cite{Hou2009} (denoted as FIGURE hereafter). The FIGURE displays the variations of the superdiffusion exponent $\alpha$ versus the screening parameter $\kappa$ for different effective coupling strength $\Gamma_\mathrm{eff}$ \cite{Kalman2004}. The authors in \cite{Comment} first claim that our observations in this figure ``resulted from an incorrect account of the coupling strength'' and instead propose a new definition of so-called relative coupling strength $\Gamma^\mathrm{rel}=\Gamma/\Gamma_{c}$, where $\Gamma_{c}$ is $\Gamma$ at melting point. However, it is not hard to prove that $\Gamma^\mathrm{rel}$ \cite{Comment} is essentially equivalent to our $\Gamma_\mathrm{eff}$ \cite{Hou2009}: $\Gamma^\mathrm{rel}=\Gamma/\Gamma_{c}=[\Gamma f(\kappa)]/[\Gamma_{c} f(\kappa)]=\Gamma_\mathrm{eff}(\kappa)/\Gamma_\mathrm{eff,c}$, where $f(\kappa)$ is a scaling formula invented by Kalman \emph{et al.} \cite{Kalman2004}. Since $\Gamma_\mathrm{eff,c}$ is a constant in \cite{Kalman2004}, $\Gamma^\mathrm{rel}$ and $\Gamma_\mathrm{eff}$ are related with a constant coefficient $\Gamma_\mathrm{eff,c}$. Second, the authors claim that the formula that we adopted in our Letter to calculate $\Gamma_\mathrm{eff}$ is only valid when $\kappa\le 3$. Unfortunately, there were no explicit or implicit statements in Ref. \cite{Kalman2004} that this formula is only valid for $\kappa\le 3$. This might be a newly-discovered issue for this formula, as we understand. In that case, it only affects the last two data points at $\kappa=3.5$ and $4.0$ in the FIGURE of \cite{Hou2009}, whereas the claim of ``incorrect account of the coupling strength'' seems exaggerated.

\begin{figure}[htp]
\centering
\includegraphics[trim=0mm 10mm 15mm 15mm,clip, width=0.48\textwidth]{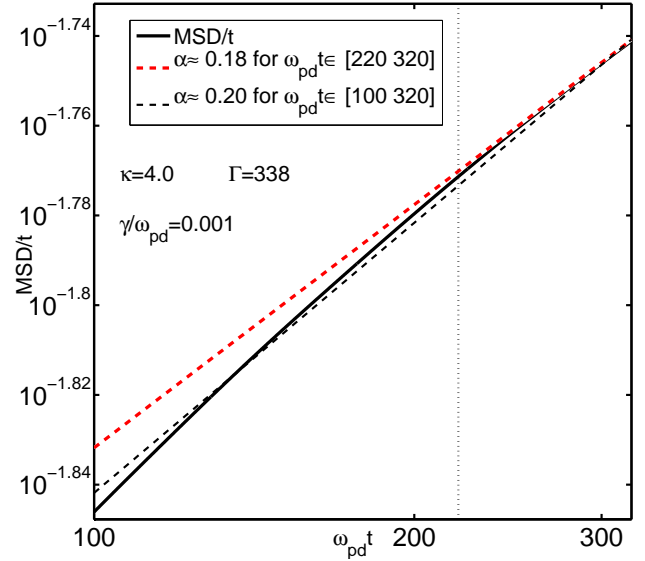}
\caption{$\text{MSD}(t)/t$ (normalized by $\omega_\mathrm{pd}a^2$) for $\kappa=4.0$,
and $\Gamma=338$ (corresponding to $\Gamma^\mathrm{rel}=0.075$ of \cite{Comment}). Dash-lines are fits of asymptotic behaviors: the upper one is fitted in the range $\omega_\mathrm{pd}t\in [220 \ 320]$ but is plotted to the full range to ease the comparison, and the lower one in the range $\omega_\mathrm{pd}t\in [100 \ 320]$. $\alpha$ is the slope of the fits.}
\label{Fig_MSD_k4_G338}
\end{figure}

In the Comment \cite{Comment} the authors also re-produce the FIGURE \cite{Hou2009} in terms of $\Gamma^\mathrm{rel}$ and make comparison between our results and theirs. We suppose that in their calculation exactly the same parameters as ours \cite{Hou2009} are used, except for the fitting range, which is $\omega_\mathrm{pd}t\in [100 \ 320]$ in \cite{Comment}. In our Letter \cite{Hou2009}, we use different fitting ranges depending on the system states (Details of fitting technique will be presented elsewhere.). In particular, for the data shown in the FIGURE, we use $\omega_\mathrm{pd}t\in [220 \ 320]$ and $\omega_\mathrm{pd}t\in [100 \ 320]$: the former gives the values of $\alpha$ and the difference of the two fits gives the uncertainties of $\alpha$. The reason for doing so is clearly shown in Fig \ref{Fig_MSD_k4_G338}, in which one observes that the tail of $\text{MSD}/t$ is not a straight line (in log-log scale). Instead its slope changes with time, so different fitting ranges will certainly result in different values of $\alpha$ (This tendency is more significant for smaller $\Gamma$ and was also observed by Donk\'{o} \emph{et al.} \cite{Donko2009} recently.). To mimimize this effect, we choose a fitting range as close to the end of the tail as possible to approximate the long time asymptotic behavior of $\text{MSD}/t$ (Note that $\alpha$ is defined by the slope of the $\text{MSD}/t$ in long time limit.), but a wider range to estimate its uncertainty. This can explain some of the differences between our results and theirs for $\kappa\le 3.0$. The dramatic differences for $\kappa=3.5$ and $4.0$ are indeed due to the different system states. We have performed new calculations with given $\Gamma^\mathrm{rel}$ in Comment \cite{Comment} and the results are shown in Table I. However, we still observe a detectable difference, especially for $\Gamma^\mathrm{rel}=0.75$: our results are substantially larger than theirs. The task of resolving all these differences is much beyond the scope of this Reply. Nevertheless, the detailed algorithms for our simulation was available in Ref. \cite{Hou2008pop} and the source code for our Letter \cite{Hou2009} will be provided upon request.

In addition, the authors of the Comment \cite{Comment} claim that the dependence of $\alpha$ on $\kappa$ is ``regular and systematic'' based on their observations. It is a very interesting extension to the results of \cite{Hou2009} and could shed new light on the effect of interaction stiffness on superdiffusion. However, the given explanation is not satisfactory and there is no any relation between their observations and the explanation.

\begin{center}
\begin{table}[htp]
\caption{$\alpha$ for $\kappa=3.5$ and $4.0$. Numerators and denominators are values of $\alpha$ fitted in the ranges $\omega_\mathrm{pd}t\in [100 \ 320]$ and $\omega_\mathrm{pd}t\in [220 \ 320]$, respectively}.
\begin{tabular*}{0.48\textwidth}{@{\extracolsep{\fill}}cccc}\hline\hline
$\kappa\backslash\Gamma^\mathrm{rel}$ &      0.075      &      0.375           &     0.75 \\
\hline
$3.5$ & $0.18/0.154$ & $0.166/0.149$   & $0.069/0.062$ \\
$4.0$ & $0.203/0.184$ & $0.164/0.156$      & $0.058/0.054$ \\
\hline\hline
\end{tabular*} \label{table1}
\end{table}
\end{center}

\end{document}